\documentclass[aps,prb,twocolumn,showpacs,amssymb,floatfix]{revtex4-1}
\usepackage{amsmath,graphics,epsfig}
\usepackage{mathrsfs}
\usepackage{bm}
\usepackage{color}

\newcommand{\tx}{\text}

\newcommand{\la}{\langle}

\newcommand{\nn}{\nonumber}

\newcommand{\ra}{\rangle}

\newcommand{\lam}{\lambda}

\newcommand{\sg}{\sigma}

\newcommand{\etal}{{\em et al.~}}
\newcommand{\ie}{{i.e.,~}}

\begin{document}
\title{Valley-dependent spin-orbit torques in two dimensional hexagonal crystals}
\author{Hang Li$^{1}$}{}
\author{Xuhui Wang$^{1}$}\email{xuhuiwangnl@gmail.com}
\author{Aur\'elien Manchon$^{1}$}\email{Aurelien.Manchon@kaust.edu.sa}
\affiliation{$^{1}$King Abdullah
University of Science and Technology (KAUST), Physical Science and
Engineering Division, Thuwal 23955-6900, Saudi Arabia}
\date{\today}

%\date{Jun10, 2014}

\begin{abstract}
We study spin-orbit torques in two dimensional hexagonal crystals such as graphene, silicene, germanene and stanene. The torque possesses two components, a field-like term due to inverse spin galvanic effect and an antidamping torque originating from Berry curvature in mixed spin-$k$ space. In the presence of staggered potential and exchange field, the valley degeneracy can be lifted and we obtain a valley-dependent Berry curvature, leading to a tunable antidamping torque by controlling the valley degree of freedom. The valley imbalance can be as high as 100\% by tuning the bias voltage or magnetization angle. These findings open new venues for the development of current-driven spin-orbit torques by structural design.
\end{abstract}

\pacs{72.25.Dc,72.20.My,75.50.Pp}
\maketitle

\section{\label{sec:intro}Introduction}

Inverse spin galvanic effect (ISGE), referring to the electrical or optical generation of a nonequilibrium spin density in non-centrosymmetric materials, has attracted much attention over the years.\cite{Ivchenko,Ivchenko-book,Vorobev,Ganichev,Yu,manchon-prb,garate-prb-2009} It originates from the momentum relaxation of carriers in an electrical field and their asymmetric redistribution in subbands that are spin-split by spin-orbit coupling.\cite{Ivchenko-book} ISGE was first observed in bulk tellurium and soon generalized to low-dimensional structures such as GaAs quantum wells.\cite{Vorobev,Ganichev,Yu}\par
 From an applied perspective, in ferromagnets lacking inversion symmetry ISGE enables the electrical control of the local magnetization through angular momentum transfer, a mechanism called spin-orbit torque (SOT).\cite{manchon-prb,garate-prb-2009} This effect has been scrutinized in dilute magnetic semiconductors such as ferromagnetic bulk (Ga,Mn)As \cite{chernyshov-nph-2009,endo-apl-2010,fang-nanotech-2011,Kurebayashi} and metallic multilayers comprising heavy metals and ferromagnets\cite{miron,liu,kim,garello,fan,jamali}. These observations have been recently extended to bilayers involving topological insulators displaying extremely large SOT efficiencies \cite{Mellnik,Fan2014}. We note that in metallic multilayers, spin Hall effect in the adjacent heavy metal also leads to a torque \cite{liu} (see discussion in Ref. \onlinecite{Brataas-nm-2012}), which complicates the interpretation of the underlying physics.\par

 From a theoretical perspective, the torque stemming from ISGE on the magnetization ${\bf M}$ has the general form 
 \begin{align}
  {\bf T}= T_{\rm DL}{\bf M}\times({\bf u}\times {\bf M})+ T_{\rm FL}{\bf M}\times{\bf u},
 \end{align}
 where the first term is called the antidamping-like torque and the second term is referred to as the field-like torque \cite{pesin2012,bijl2012,Hang-apl-2013,Hang-2015,wang-manchon-2012,Lee2015}. The antidamping-like torque is {\em even} in magnetization direction and competes with the antidamping, while the field-like torque is {\em odd} in magnetization direction and acts like a magnetic field. The vector ${\bf u}$ depends on the current direction {\bf j} and the symmetries of the spin-orbit coupling. For instance, in a ferromagnetic two-dimensional electron gas (normal to ${\bf z}$) with Rashba spin-orbit coupling, ${\bf u}={\bf z}\times{\bf j}$.\cite{manchon-prb} An interesting aspect of the formula given above is that the antidamping-like torque arises from the distortion of the wavefunction induced by the electric field, a mechanism closely related to the material$'$s Berry curvature \cite{Kurebayashi,bijl2012,Hang-2015,Lee2015}.\par

In parallel to the development of SOT in ferromagnetic structures, the study of spin-orbit coupled transport has also been extended to low-dimensional hexagonal crystals such as graphene. Experimentally, a spin-splitting induced by Rashba spin-orbit coupling has been observed in graphene grown on heavy metals or surface alloys.\cite{Varykhalov, Marchenko, Leicht} Furthermore, a ferromagnetic insulator EuO was successfully deposited on graphene and spin-polarized states were detected.\cite{Swartz,Klinkhammer,Klinkhammer2} The recent fabrication of low-dimensional hexagonal crystals with strong intrinsic spin-orbit coupling such as silicene \cite{Fleurence2012,Vogt-2012}, germanene \cite{Li-2014} and possibly stanene \cite{Xu-2013}, has enriched the graphene physics. These materials offer a rich platform for the investigation of spin, orbital and valley-dependent phenomena\cite{Han-2014,Manchon-2015}.\par

In this paper, we theoretically investigate the nature of SOT in two-dimensional hexagonal IV group elements crystals such as graphene, silicene, germanene and stanene. As a matter of fact, the wide tunability of their model band structure presents an appealing opportunity to study the impact of the band geometry (e.g. their Berry curvature) on nonequilibrium mechanisms. Using Kubo formula, we investigate the impact of the band structure on the different components of SOT. We find that intrinsic spin-orbit coupling affects the antidamping-like and field-like components differently. The former is sensitive to the presence of a staggered potential while the latter is not. We understand these results in terms of Berry curvature origin of the antidamping torque. The presence of both magnetization and staggered potential enables the emergence of a valley-dependent antidamping torque, providing an additional degree of freedom to the system. 
                    
%---------- model ----------

\section{\label{sec:model}Model and Method}
\begin{figure}[tbh]
\begin{center}
\includegraphics[trim=0mm 0mm 0mm 0mm,clip,scale=0.4]{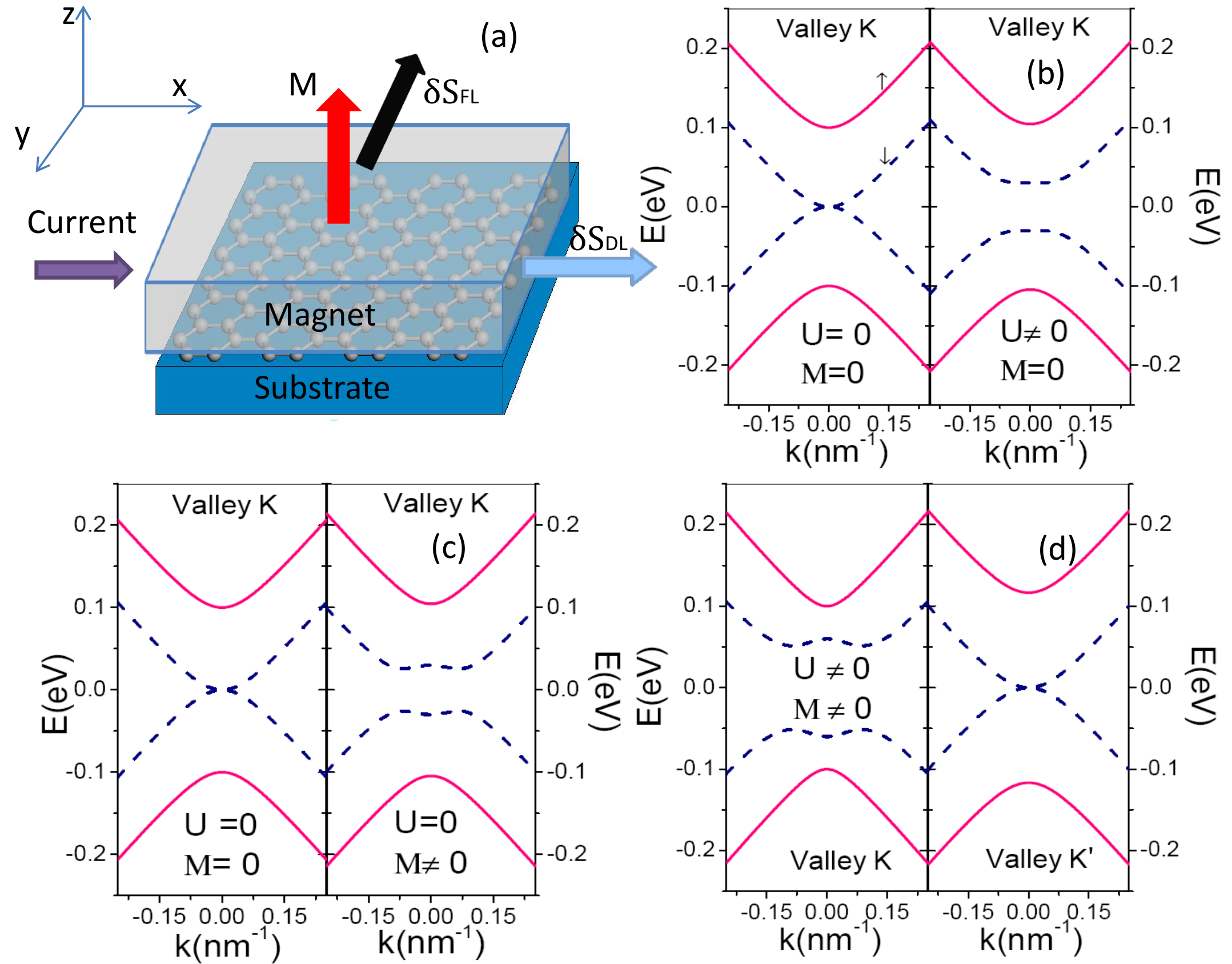}
\end{center}
\caption{(Color online) Schematics of the device based on graphene-like materials with field-like and antidamping-like SOT. (b)-(d) Energy dispersion of graphene-like materials with (b) $U = 0.03~eV$ (c) $M=0.03~eV$ (d) $U = 0.03~eV$ and $M= 0.03~eV$. The current flows from left to right. Magnetization is assumed to be directed along the z-axis.}
\label{fig1}
\end{figure}

A possible structure to realize valley-dependent SOT is a single-layered hexagonal lattice (such as graphene, silicene, germanene or stanene) sandwiched by a ferromagnetic layer and a non-magnetic substrate [see Fig. 1(a)]. The ferromagnetic layer may be chosen as EuO,\cite{Swartz} or YIG,\cite{Wang} which induces a weak exchange coupling on the spin-polarized carriers. The underlying non-magnetic substrate provides Rashba spin-orbit coupling.\cite{Marchenko,Leicht,Jin,Chang} Note that in principle, a magnetic insulator could supply for both exchange field and Rashba spin-orbit coupling.\cite{Fan2014,Zou,Qiao2}\par

The concept of valley-dependent SOT is illustrated in Fig. \ref{fig1}(a). In the absence of a magnet, the interaction between the substrate and graphene-like layer breaks the inversion symmetry and leads to a Rashba spin-orbit coupling. As a results, a transverse nonequilibrium spin density builds up when a current is injected along the horizontal direction. Both Rashba and intrinsic spin-orbit coupling are valley dependent as shown in Eq. (\ref{eq:total-Hamiltonian}) and thus they can not break the valley degeneracy. In the presence of a magnet, a field-like spin density and an antidamping-like spin density are generated as shown in Fig. \ref{fig1}(a).\cite{miron,liu} The exchange field only breaks the time-reversal-symmetry while the sublattice symmetry (two-fold rotational symmetry in the plane) is preserved as shown in Fig. \ref{fig1}(c). The interaction between the substrate and graphene-like layer can also induce a staggered potential, which enlarges the band gap without affecting the valley degeneracy, as shown in Fig. \ref{fig1}(b).\par
 However, in the presence of both staggered potential and ferromagnetic exchange field, the valley degeneracy is lifted since both the time-reversal and sublattice symmetries are broken as shown in Fig. \ref{fig1}(d). As a result, SOT becomes valley dependent. Furthermore, as discussed in the next section, the band structure distortion displayed in Figs. \ref{fig1}(b)-(d) affects the magnitude of the SOT components.\par
 
   We adopt a low-energy continuum model Hamiltonian which describes Dirac electrons near to the Fermi energy and captures the physics behind the formation of the valley-dependent SOT in the vicinity of K and K$'$ points. The total Hamiltonian at K or K$'$ valley in the basis of $\{\psi_{A,\uparrow},\psi_{B,\downarrow},\psi_{B,\uparrow},\psi_{A,\downarrow}\}$ reads\cite{Qiao}
\begin{align}
H_{\tx{sys}}&=v(\tau k_{x}\hat\sg_{x}- k_{y}\hat\sg_{y})\otimes\hat{\rm I}+\frac{\lam_{R}}{2}(\tau \hat\sg_{x}\otimes\hat s_{y}-\hat\sg_{y}\otimes \hat s_{x})\nn\\
&+\tau \lam_{so}\hat\sg_{z}\otimes\hat  s_{z}+J_{\rm ex} \hat{\rm I} \otimes {\bf M}\cdot\hat{\bf s}+U\hat\sg_{z}\otimes\hat{\rm I},
\label{eq:total-Hamiltonian}
\end{align} 
 where $v=\sqrt{3}at/2$ with t being a nearest-neighbor hopping parameter, $\tau=+1(-1)$ stands for the K or (K$'$) valley, $\hat{\rm I}$ is a 2$\times$ 2 unity matrix,  $a$ is the lattice constant and $J_{\rm ex}$ is the ferromagnetic coupling constant. $\hat {\bm \sg}$ and $\hat {\bf s}$ are Pauli matrices denoting the AB-sublattice and spin degrees of freedom, respectively. ${\bf M}$ is the magnetization direction. The first term includes the spin-independent kinetic energy of the particle, the second term denotes the Rashba coupling and the third one represents the intrinsic spin-orbit coupling. The fourth term is the interaction between the spin of the carrier and the local moment of the ferromagnetic system. The last term is the staggered potential (induced, for instance, by an electrical field or a substrate \cite{Ezawa2012,Kane,Zhou2007}), where $U=1~(-1)$ for A (B) site.\par
    To compute the current-induced effective magnetic field, we first evaluate the nonequilibrium spin density $\delta {\bf S}$ at K (K$'$) valley using Kubo formula:\cite{Kurebayashi}
\begin{align}
\delta{\bf S}_{K(K')}=&\frac{e\hbar}{2\pi V}{\rm Re}\sum_{{\bf k},a,b} \la\psi_{{\bf k}b}|\hat{\bf{s}}|\psi_{{\bf k}a}\ra\la\psi_{{\bf k}a}|  {\bf E}\cdot \hat{\bf{v}}|~\psi_{{\bf k}b}\ra\nn\\
&\times [G^{R}_{{\bf k}b}G^{A}_{{\bf k}a}-G^{R}_{{\bf k}b}G^{R}_{{\bf k}a}],
\label{eq:TSP}
\end{align}
where {\bf E} is the electric field, $\hat{\bf{v}}=\frac{1}{\hbar}\frac{\partial H}{\partial {\bf{k}}}$ is the velocity operator, $G^{R}_{{\bf k}a}=(G^{A}_{{\bf k}a})^{*}=1/(E_{F}-E_{{\bf k}a}+i\Gamma)$. $\Gamma = \hbar / 2\tau$ is the disorder-induced energy spectral broadening due to the finite life time of the particle in the presence of impurities and $\tau$ is the momentum scattering time. $E_{F}$ is the Fermi energy, $E_{{\bf k}a}$ is the energy of electrons in band a. The eigenvector $|\psi_{{\bf k},a}\ra$ in band $a$ can be found by diagonalizing Eq. (\ref{eq:total-Hamiltonian}). Equation (\ref{eq:total-Hamiltonian}) contains both intraband ($a=b$) and interband ($a\neq b$) contributions to the nonequilibrium spin density. Simpler expressions in the weak $\Gamma$ limit can be found in Ref. \onlinecite{Hang-2015}. The former stems from the perturbation of the carrier distribution function by the electric field and it is inversely proportional to $\Gamma$. The latter arises from the perturbation of the carrier wave functions by the electric field. The interband contribution also depends on $\Gamma$ but survives when $\Gamma \rightarrow 0$.

 In order to evaluate the current-driven SOT in different materials, we define the electrical efficiency of the torque as \cite{manchon-prb}
\begin{align}\label{eq:efficiency of density}
\eta=\frac{2J_{\rm ex}\delta{S}}{\hbar\sg_{xx}{\rm E}}
\end{align}
where $\sg_{ij}$ is conductivity tensor component defined\cite{Murayama}
\begin{align}\label{eq:conductivity}
\sg_{ij}&={e^2\hbar}{\rm Re}\sum_{{\bf k},a,b}
[{\la\psi_{{\bf k}a}| {\bf{\hat v}}_{i}|~\psi_{{\bf k}b}\ra}{\la\psi_{{\bf k}b}| {\bf{\hat v}}_{j}|~\psi_{{\bf k}a}\ra}]\nn\\
&\times  [G^{R}_{{\bf k}b}G^{A}_{{\bf k}a}-G^{R}_{{\bf k}b}G^{R}_{{\bf k}a}].
\end{align}

\section{\label{sec:numerical results}Inverse spin Galvanic effect}
The characteristics of the SOT in two-dimensional hexagonal honeycomb lattices are expected to be different from the well studied case of bulk GaMnAs\cite{manchon-prb,garate-prb-2009,pesin2012,bijl2012,Hang-apl-2013,Hang-2015}. Unlike the three-dimensional ferromagnetic GaMnAs in the weak limit $(\lambda_R\ll J_{\rm ex},~J_{\rm ex}\sim 1eV \rm~and~\lambda_R\sim 0.1eV)$, the graphene-like materials often fall into the strong limit $(\lambda_R\gg J_{\rm ex})$, leading to a nonzero interband contribution. The nontrivial Dirac kinetic term [first term in Eq. (\ref{eq:total-Hamiltonian})] gives rise to nonlinear transitions of spin density when tuning the Fermi energy. Furthermore, the spin density is more sensitive to band topology tunable by intrinsic spin-orbit coupling or staggered potential. More importantly, the Dirac kinetic term and spin-orbit coupling terms are valley-dependent. In order to better understand the valley-dependent SOT, we first examine spin torque with and without valley degeneracy in section~\ref{sec:numerical results} and \ref{sec:numerical results2} respectively.

\subsection{Non-magnetic honeycomb lattice}
\begin{figure}[tbh]
\begin{center}
\includegraphics[trim=0mm 0mm 0mm 0mm,clip,scale=0.4]{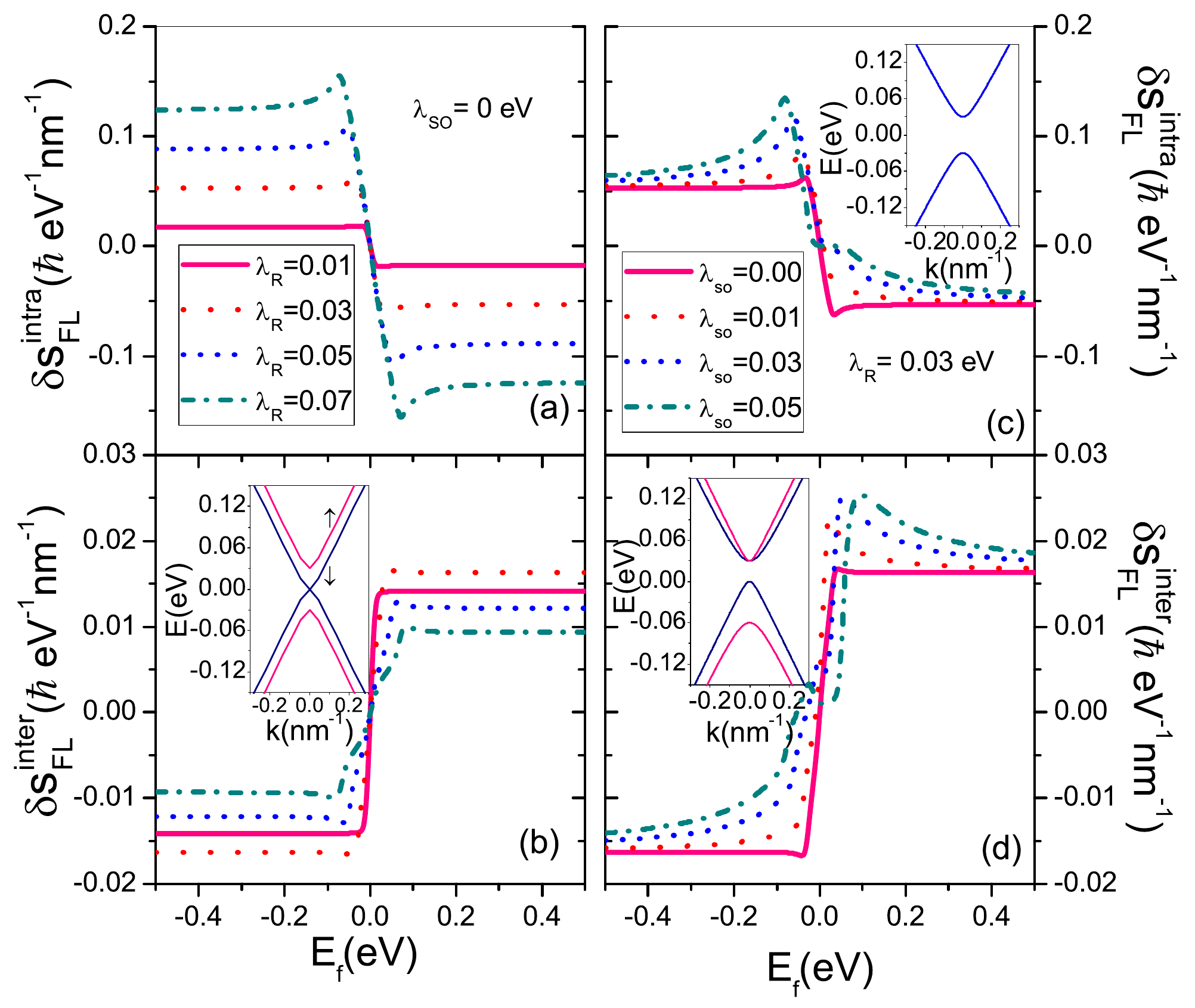}
\end{center}
\caption{(Color online) (a) Intraband and (b) interband contributions to spin density as a function of Fermi energy $E_{f}$ for various Rashba spin-orbit coupling in the absence of intrinsic spin-orbit coupling $\lam_{so}$. (c) Intraband and (d) interband spin density as a function of Fermi energy $E_{f}$ for various intrinsic spin-orbit coupling at $\lam_{R}$=0.03 eV. Inset (b): Band structure of graphene-like materials with $\lam_{R}$=0.03 eV and $\lam_{so}$=0 eV. Inset (c): Band structure with $\lam_{R}$=0 eV and $\lam_{so}$=0.03 eV Inset (d): Same as inset (b) but with $\lam_{so}$=0.03 eV. The current is injected along the x axis.}
\label{fig2}
\end{figure}

We first compute the spin density induced by ISGE in {\em non-magnetic} graphene. In this material, we choose the following parameters: $E_f\in$ [0, 0.3] eV,\cite{Pesin} $\lambda_R\in$  [10, 130] meV,\cite{Marchenko,Leicht} and $J_{\rm ex}\in$  [5, 30] meV.\cite{Haugen,Yang1} For all the calculations shown in this section, the electrical field is assumed to be along the $x$-axis and the energy broadening is $\Gamma=$0.01 eV. To understand the physical origin of the SOT and establish connections with previous works (such as Ref. \onlinecite{Hang-2015}), we parse the SOT into intraband and interband contributions.\par

Figure \ref{fig2} presents the intraband (a,c) and interband contributions (b,d) to the ISGE-driven spin density for various strengths of $\lambda_{\rm R}$ (a,b) and $\lambda_{\rm so}$ (c,d). In non-magnetic graphene, intraband contribution produces a spin density aligned toward the $y$-direction, which is expected from the geometry of our system and consistent with the well known ISGE in two-dimensional electron gases \cite{edelstein,manchon-prb}. There is also a non-ignorable interband contribution in the strong limit $(\lambda_R\gg J_{\rm ex})$, smaller than the intraband contribution and opposite to it, in agreement with our previous analytical solutions in the case of Rashba two-dimensional electron gas\cite{Hang-2015}. When increasing the absolute value of Fermi energy, the spin density first experiences a sharp enhancement at small values of $E_{f}$ and quickly saturates. This result is consistent with Ref. \onlinecite{Dyrdal} and can be readily understood by considering the band structure in the inset of Fig. \ref{fig2}(b). When the Fermi energy lies in the energy gap of two spin-split subbands, only one spin species contributes to ISGE and the intraband spin density increases with the Fermi energy. As the Fermi energy lies above the subband gap, the two subbands compensate each other and the spin density saturates. The peaks in Fig. \ref{fig2}(a) correspond to the minimum ($E > 0$) or maximum ($E < 0$) of the spin-up subband [see inset of Fig. \ref{fig2}(b)] which is of the order of $\lambda_{\rm R}$. \par

Another interesting feature is the spin density as a function of the Rashba spin-orbit coupling. The intraband contribution increases linearly with $\lambda_{\rm R}$ [see Fig. \ref{fig2}(a)], while the interband contribution first increases and then decreases [see Fig. \ref{fig2}(b)]. The interband contribution depends on the energy difference between the subbands, which itself is of the order of $\lambda_{\rm R}$. Indeed, one can show that in the weak impurity limit, the interband contribution is proportional to 1/($ E_{{\bf k}a}$-$ E_{{\bf k}b}$).\cite{Kurebayashi,Hang-2015,Lee2015} This results in the non-linear dependence as a function of $\lambda_{\rm R}$ observed in Fig. \ref{fig2}(b) as well as in Fig. 3(c). \par

Rashba spin-orbit coupling is not the only spin-orbit coupling that affects the spin density. In graphene-like systems Rashba spin-orbit coupling is always accompanied by an intrinsic spin-orbit coupling, $\sim \tau \lam_{so}\hat\sg_{z}\otimes\hat  s_{z}$, which originates from the substrate or a low buckled structure.\cite{Ezawa2012,Liu-2011}In Figs. \ref{fig2}(c) and (d), we display the Fermi energy dependence of the intraband and interband contributions to spin density for various intrinsic spin-orbit coupling. As expected, the intrinsic spin-orbit coupling opens up a band gap and distorts the topology of the band structure as seen in the inset of Fig. \ref{fig2}(c) and (d). For a given K or K$'$ Valley (ignore $\tau$), this term plays the same role as the ferromagnetic exchange field along the z axis in unit cell when the two sublattices contribute to spin density equivalently ($\sg_{z}$ replaced by $\hat{\rm I}$ ). When the two sublattices contribute to spin density inversely, this term acts as an anti-ferromagnetic exchange field and the symmetry of profiles of the spin density is broken and it shifts to the left. Furthermore, the asymmetry of the profiles of the spin density becomes more evident with the increase of $\lam_{so}$. The energy at which the spin density is maximum equals $\lam_{so}+\lam_{R}$ when $ E_f$ $<$ 0. Note that the intrinsic spin-orbit coupling does not drive ISGE by itself, but it affects the ISGE-induced spin density driven by Rashba spin-orbit coupling through the modulation of the topology of the bands.

\subsection{Magnetic honeycomb lattice}

\begin{figure}[tbh]
\begin{center}
\includegraphics[trim=0mm 0mm 0mm 0mm,clip,scale=0.3]{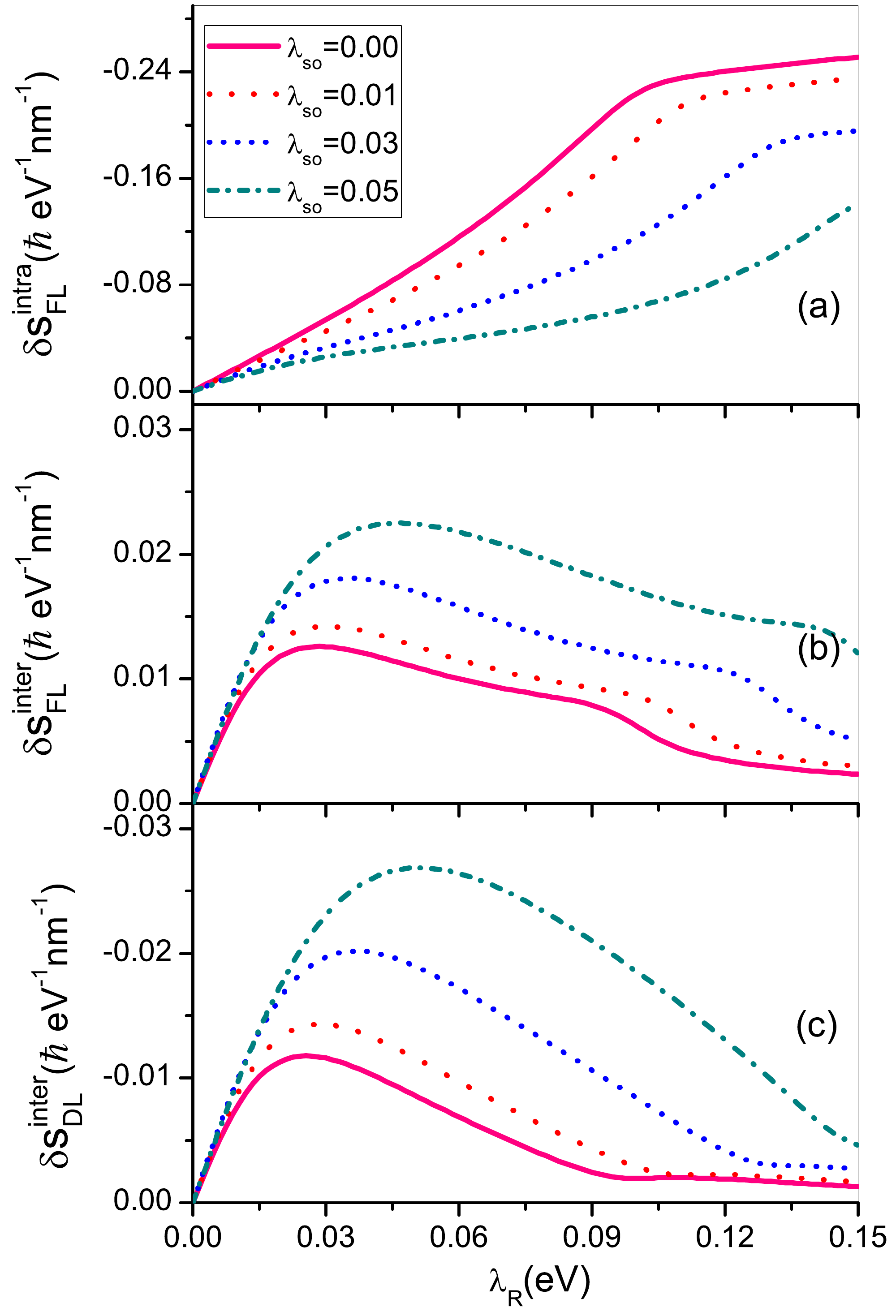}
\end{center}
\caption{(Color online)(a) Intraband and (b)-(c) interband spin density as a function of Rashba spin-orbit coupling for different intrinsic spin-orbit coupling with $J_{\rm ex}=0.01~eV$ and $E_f=0.1~eV$. The magnetization is directed along the z axis.}
\label{fig3}
\end{figure}

Let us now turn to the case of {\em magnetic} two-dimensional honeycomb lattices. To understand the role of spin-orbit coupling, we plot the intraband and interband spin density as a function of Rashba spin-orbit coupling for different intrinsic spin-orbit coupling in the presence of magnetization in Fig. \ref{fig3}. Due to the presence of magnetism, the interband contribution also produces an antidamping component [see Figs. \ref{fig3}(c)], i.e. a spin density contribution oriented towards $\sim{\bf M}\times{\bf y}$ \cite{Kurebayashi,Hang-2015,Lee2015} and with a magnitude comparable to the one of the field-like component [see Figs. \ref{fig3}(c)]. As seen in Figs. \ref{fig3}(a)-(c) the interband field-like and antidamping contributions first increase and then decrease. This can be understood as a competition between the spin density driven by Rashba spin-orbit coupling and the suppression of interband scattering due to the distance between the subbands that increases with $\lambda_{\rm R}$.\par

The intraband contribution decreases with the increasing intrinsic spin-orbit coupling while the interband contribution behaves the opposite way. By opening a band gap, the intrinsic spin-orbit coupling alters the band filling, resulting in a reduced intraband contribution to spin density. An analytical solution of energy depending on intrinsic spin-orbit coupling can be found in Ref. \onlinecite{Qiao3}. On the other hand, the intrinsic spin-orbit coupling reduces the splitting between the subbands for $ E_{f}>0$ [see inset in Fig. \ref{fig2}(d)], which results in an enhancement of the interband contributions. This result is valuable to current-driven magnetic excitations since the antidamping torque is responsible for magnetization switching and excitations\cite{miron,liu} (see also, for instance, Ref. \onlinecite{demidov}).

%%%%%%%%%%%%%%%%%%%%%%%%%%%%%%%%%%%%%%%%%%%%%%%%%%
\section{\label{sec:numerical results2}Valley-Dependent Spin-Orbit Torque}
The valley degree of freedom can be used as a tool to enhance the functionality of two-dimensional honeycomb lattices.\cite{Rycerz} Recently, a valley-dependent anomalous quantum hall state has been predicted in silicene and silicene nanoribbons owing to the topological phase transition \cite{Ezawa2012,Pan2014}. A charge-neutral Hall effect has been measured in graphene devices\cite{Gorbachev2014,Abanin2011}. These suggest the emergence of valley Hall effect. It is thus natural to expect a valley-modulated SOT in our settings.

\subsection{Staggered Potential}

\begin{figure}[tbh]
\centering
\includegraphics[trim = 0mm 0mm 0mm 0mm, clip, scale=0.4]{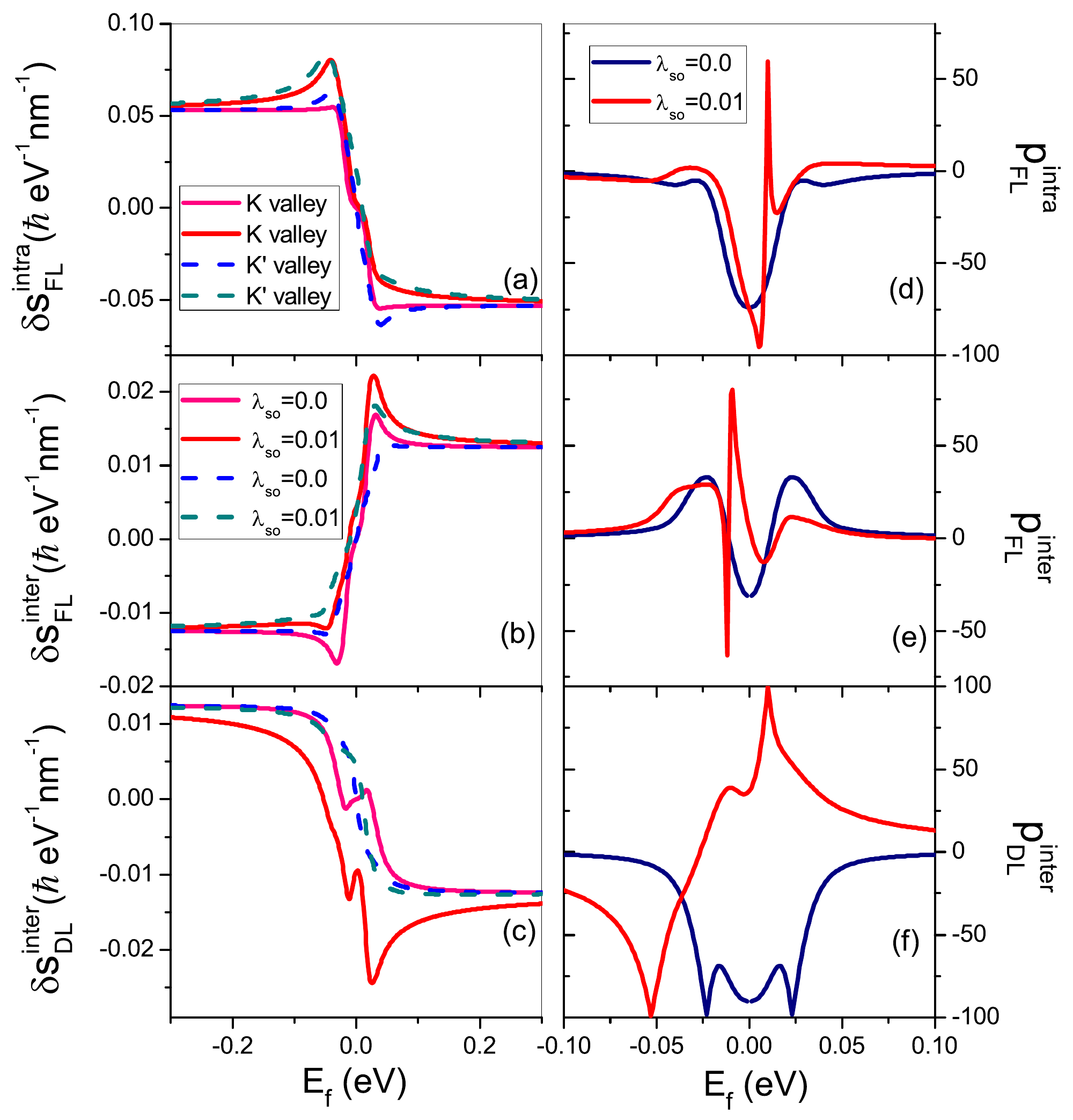}
\caption{(Color online) (a) Intraband and (b)-(c) interband spin density of two valleys as a function of Fermi energy for different intrinsic spin-orbit coupling with $U=0.01~eV$ and $J_{\rm ex}=0.01~eV$. Valley polarization for intraband  (d) and interband (e)-(f) components for different intrinsic spin-orbit coupling.
}
\label{fig4}
\end{figure}

The sublattice degeneracy can be removed by depositing graphene-like materials on hexagonal boron-nitride\cite{Gorbachev2014,Min-2008,Hunt2013} or silicon carbide,\cite{Zhou2007} or by applying an electrical field in a low buckled structure \cite{Ezawa2012}. When the staggered potential and exchange field are present and the valley degeneracy is broken, the spin density becomes valley-dependent as shown in Figs. \ref{fig4}.\par

  In Figs. \ref{fig4}(a)-(c), we display the intraband and interband contributions to spin density as a function of Fermi energy in the presence of staggered potential with and without the intrinsic spin-orbit coupling. The imbalance between the contribution of the two valleys to the spin density, \ie valley polarization, defined as $P=\frac{\delta{\bf S}_{K}-\delta{\bf S}_{K'}}{\delta{\bf S}_{K}+\delta{\bf S}_{K'}}$, is reported on Figs.\ref{fig4}(d)-(f). The largest imbalance occurs mainly around the neutrality point $ E_f = 0$. The valley imbalance of the antidamping-like component can reach 100\% as shown in Figs.\ref{fig4}(f), \ie that for certain energies, this component is dominated by only one valley. When the intrinsic spin-orbit coupling is present, the magnitudes of the valley imbalance can be switched from -100\% to 100\% by simply tuning the Fermi energy.\par

\begin{figure}[tbh]
\centering
\includegraphics[trim = 0mm 0mm 0mm 0mm, clip, scale=0.4]{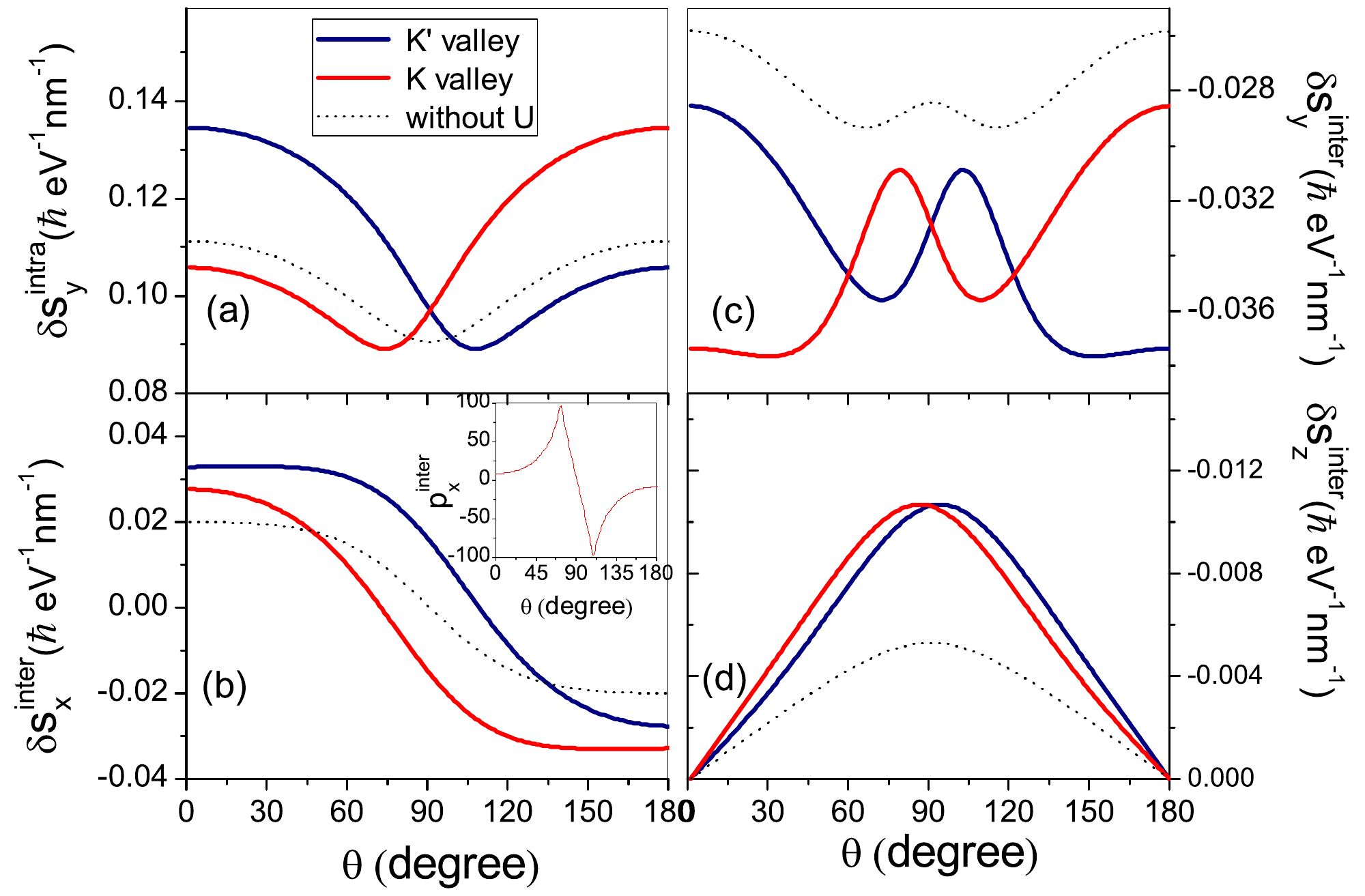}
\caption{(Color online) Intraband and interband spin density as a function of the
magnetization direction with (solid lines) and without (dashed lines)staggered potential for the different valleys when U = 0.03 eV. Inset (b) Valley polarization of interband spin density for x-component. 
}
\label{fig5}
\end{figure}

\subsection{Angular dependence}
A noticeable effect of lifting the valley degeneracy is its impact on the angular dependence of SOT components. Figure \ref{fig5} displays the angular dependence of the different components of the spin density when the magnetization is rotated in the (x,z) plane. In a ferromagnetic two-dimensional electrons gas with Rashba spin-orbit coupling, the spin density has the general form $\delta {\bf S}=\delta S_\|\cos\theta{\bf x}+\delta S_y{\bf m}\times{\bf y}-\sin\theta\delta S_\|{\bf z}$ (e.g., see Ref. \onlinecite{Hang-2015}), where $\theta$ is the angle between the magnetization and {\bf z}. More complex angular dependence may appear in the strong Rashba limit $(\lambda_R\gg J_{\rm ex})$ due to D$'$yakonov-Perel relaxation \cite{Ortiz2013} or in the intermediate regime $(\lambda_R \sim J_{\rm ex})$ due to a "breathing" Fermi surface \cite{Lee2015}. \par
   Similarly, in the case of magnetic honeycomb lattices, different components of the spin density display a clear deviation from the simple $\sim\cos\theta$ dependence of the ferromagnetic Rashba gas (see dotted lines in Fig. \ref{fig5}). This is attributed to the "breathing" Fermi surface, \ie the distortion of the Fermi surface, and the modification of the band filling as a function of the direction of the magnetization when the exchange is comparable to the Rashba parameters.\par

In the absence of valley degeneracy, the angular dependence at K and K' points differ significantly from each other (red and blue lines in  Fig. \ref{fig5}, respectively). As a consequence, by tuning the magnetization angle the valley imbalance varies strongly [from -$100\%$ to $100\%$ for the x-component, as shown in inset of in Fig. \ref{fig5}(b)]. We also notice that additional structures are visible in the angular dependence of the field-like component, related to interband transitions [see Fig. \ref{fig5}(c)]. These features are unique to the case of honeycomb lattices and absent in standard two dimensional free electron gases.

\begin{figure}[tbh]
\centering
\includegraphics[trim = 0mm 0mm 0mm 0mm, clip, scale=0.4]{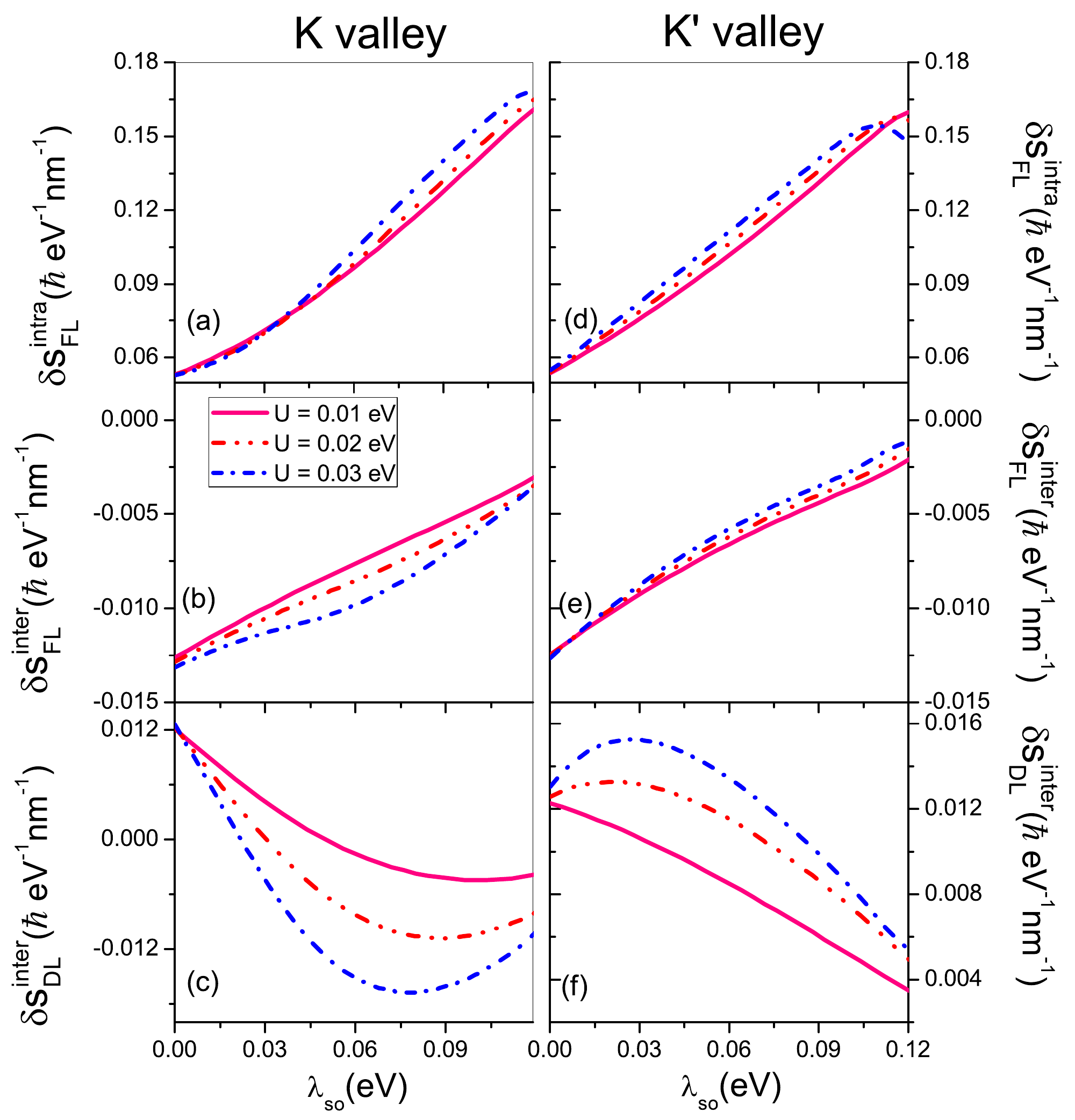}
\caption{(Color online) Intraband and interband spin density as a function of intrinsic spin-orbit coupling for different staggered potential for the K valley (a)-(c) and K$'$ valley (d)-(f). The parameters are: $E_f = -0.16~eV$, and $J_{\rm ex}=0.01~eV$ and $\lambda_R=0.03~eV$. 
}
\label{fig6}
\end{figure}

\begin{figure}[tbh]
\centering
\includegraphics[trim = 0mm 0mm 0mm 0mm, clip, scale=0.4]{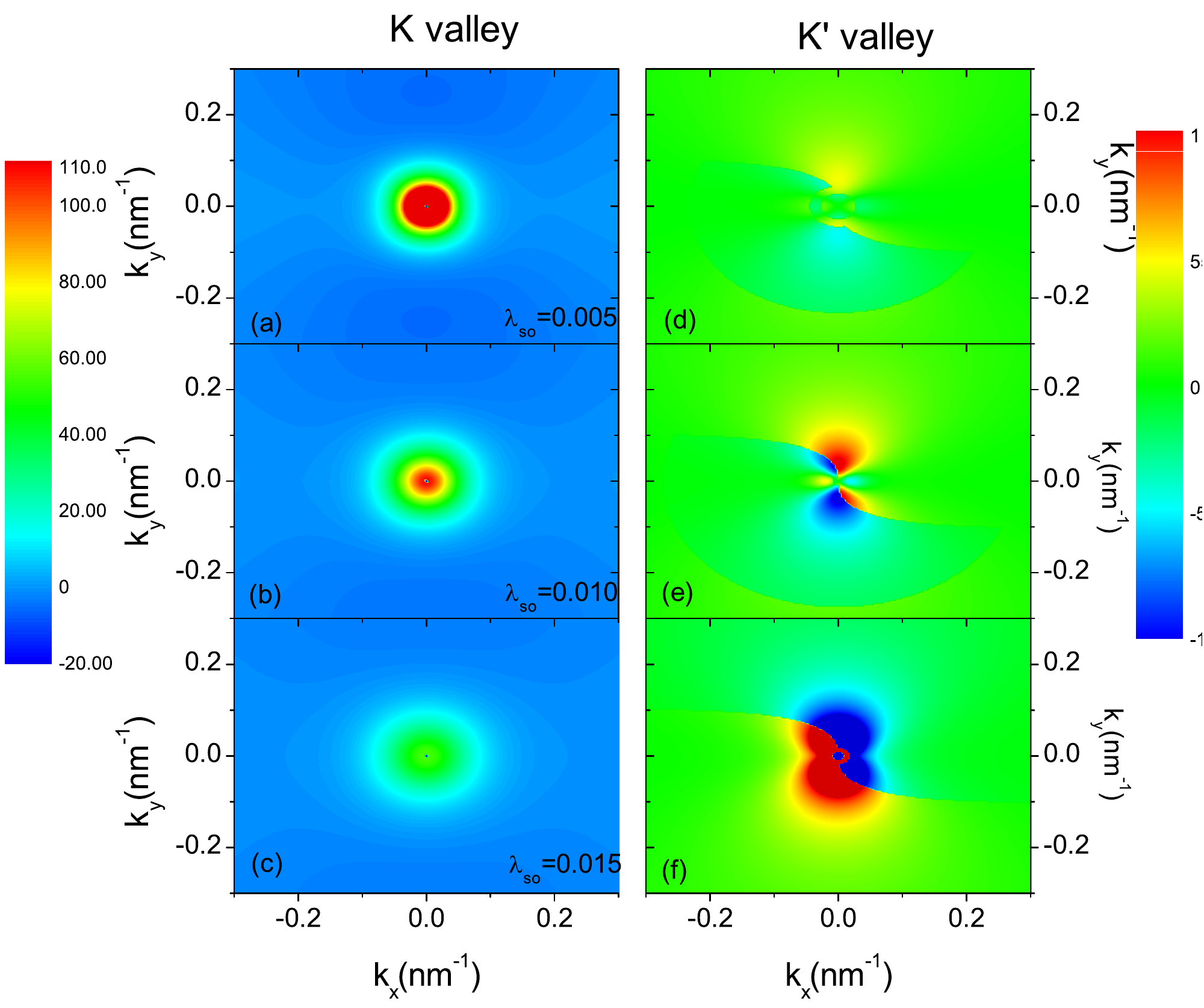}
\caption{(Color online) Contour of valley-polarized Berry curvature distribution for different intrinsic spin-orbit coupling in $k_{x}-k_{y}$ plane with U = 0.03 eV. (a) and (d) $\lam_{so}$ = 0.005~eV. (b) and (e) $\lam_{so}$ = 0.01~eV. (c) and (f) $\lam_{so}$ = 0.015~eV.  Others parameters are the same as in Fig.\ref{fig6}. 
}
\label{fig7}
\end{figure}

\section{Connection Between Spin-Orbit Torque and Berry curvature}
  Berry's phase plays a crucial role in the transport properties of semiconductors especially for graphene-like materials. Due to the inequivalent contribution from two valleys, Berry curvature induces valley hall effect in graphene with broken inversion symmetry.\cite{Gorbachev2014} Recently, the link between SOT and Berry curvature was established in bulk ferromagnetic GaMnAs.\cite{Kurebayashi} The intrinsic spin-orbit coupling distorts the Fermi surface and gives rise to the oscillations in torque magnitudes, as already observed in (Ga,Mn)As\cite{Kurebayashi}.\par

In order to show the connection between the SOT and the band structure distortion, let us analyze the influence of intrinsic spin-orbit coupling on SOT in the presence of a staggered potential. The intraband and interband contributions to spin density as a function of intrinsic spin-orbit coupling for various staggered potential both at K and K$'$ valley are displayed in Fig. \ref{fig6}. We find that both the field-like intraband and interband contributions to the spin density, $\delta S_{\rm FL}^{\rm intra}$ and $\delta S_{\rm FL}^{\rm inter}$. They increase with the intrinsic spin-orbit coupling and are only weakly affected by the staggered potential [see Figs. \ref{fig6} (a,b) and (d,e)]. In contrast, the antidamping-like component of the spin density, $\delta S_{\rm DL}^{\rm inter}$, displays a non-linear dependence as a function of the intrinsic spin-orbit coupling that is very different for the two valleys and highly sensitive to the staggered potential [see Figs. \ref{fig6} (c) and (f)].\par

 To understand this difference, we plot the contour of Berry curvature for different intrinsic spin-orbit coupling at K and K$'$ valleys in $k_{x}-k_{y}$ plane in Fig. \ref{fig7}. A large Berry curvature mainly concentrates around the Dirac point and decays away from it, in agreement with previous results\cite{Xiao,Qiao}. For the K valley, Berry curvature decreases with the increase of intrinsic spin-orbit coupling. Yet for the K$'$ valley, Berry curvature increases. This trend is in accordance with the variations of $\delta S_{\rm DL}^{\rm inter}$ displayed in Figs. \ref{fig6} (c) and (f) and not in accordance with the variations of $\delta S_{\rm FL}^{\rm inter}$ displayed in Figs. \ref{fig6} (b) and (e). It illustrates the fact that while $\delta S_{\rm FL}^{\rm inter}$ and $\delta S_{\rm DL}^{\rm inter}$ both originate from interband transitions, only the latter is related to Berry curvature, \ie the field-like SOT is purely due to ISGE instead of the superposition of Berry curvature and ISGE in ferromagnetic GaMnAs as pointed out by Kurebayashi et al.\cite{Kurebayashi}
 
 \section{Discussion}

To complete the present study, we computed the magnitude of antidamping-like and field-like components of the spin density and corresponding electrical efficiencies for various graphene-like honeycomb lattices, assuming $\lam_{R}=0.1$ eV and $J_{\rm ex}=0.03$ eV. The results are reported in Table 1, showing that the largest SOT is obtained for stanene [$\sim 100\times10^{10}$  $\rm eV/(A\cdot m)$]. As a comparison, the corresponding efficiencies of field-like SOT in (Ga,Mn)As\cite{Hang-2015}, two-dimensional Rashba systems\cite{manchon-prb} and topological insulators\cite{Sakai} are of the order of $\sim 1\times10^{10}$, $\sim 10\times10^{10}$ and $\sim 100\times10^{10}$  $\rm eV/(A\cdot m)$, respectively, in agreement with the orders of experimental results\cite{chernyshov-nph-2009,fang-nanotech-2011,miron,Fan2014}. Therefore, for moderate Rashba and exchange parameters, honeycomb lattices seem to display large field-like torques. Interestingly, the antidamping-like torque remains about one order of magnitude smaller than the field-like torque, as already observed in two-dimensional Rashba gases and (Ga,Mn)As \cite{Hang-2015}.

\begin{table*}
\renewcommand{\arraystretch}{1.1}
\caption{Efficiency of spin torque for various two dimensional hexagonal lattices} \label{table_1} \centering
\begin{tabular}{||c|c|c|c|c|c|c|c||} 
\hline     
\hline    
$$&$\tt{E}(eV)$&$\tt{a}(\AA)$&$\tt{\sg_{xx}}(e^{2}/\hbar)$&$s_{DL}(\hbar({eV nm})^{-1})$&$s_{FL}(\hbar({eV nm})^{-1})$&$\eta_{DL}(eV {(A m)}^{-1})$&$\eta_{FL}(eV {(A m)}^{-1})$\\   
\hline      
Carbon&$2.7^{\cite{Qiao}}$&$2.46^{\cite{Liu-2011}}$&$23.3809\times10^{-3}$&$0.0083$&$0.1193$&$2.13\times10^{10}$&$30.6\times10^{10}$\\       
\hline
 Silicene&$1.04^{\cite{Tsai-2013}}$&$3.87^{\cite{Liu-2011}}$&$9.0068\times10^{-3}$&$0.0137$&$0.1975$&$9.13\times10^{10}$&$131.6\times10^{10}$\\      
\hline
 Germanene&$0.97^{\cite{Tsai-2013}}$&$4.06^{\cite{Liu-2011}}$&$8.4004\times10^{-3}$&$0.0141$&$0.2019$&$10.07\times10^{10}$&$144.2\times10^{10}$\\
 \hline
 Stanene&$0.76^{\cite{Tsai-2013}}$&$4.67^{\cite{Liu-2011}}$&$6.5818\times10^{-3}$&$0.0155$&$0.220$&$14.13\times10^{10}$&$200.6\times10^{10}$\\    
 \hline     
 \hline     
\end{tabular}\\
\end{table*}
  
\begin{figure}[tbh]
\centering
\includegraphics[trim = 0mm 0mm 0mm 0mm, clip, scale=0.4]{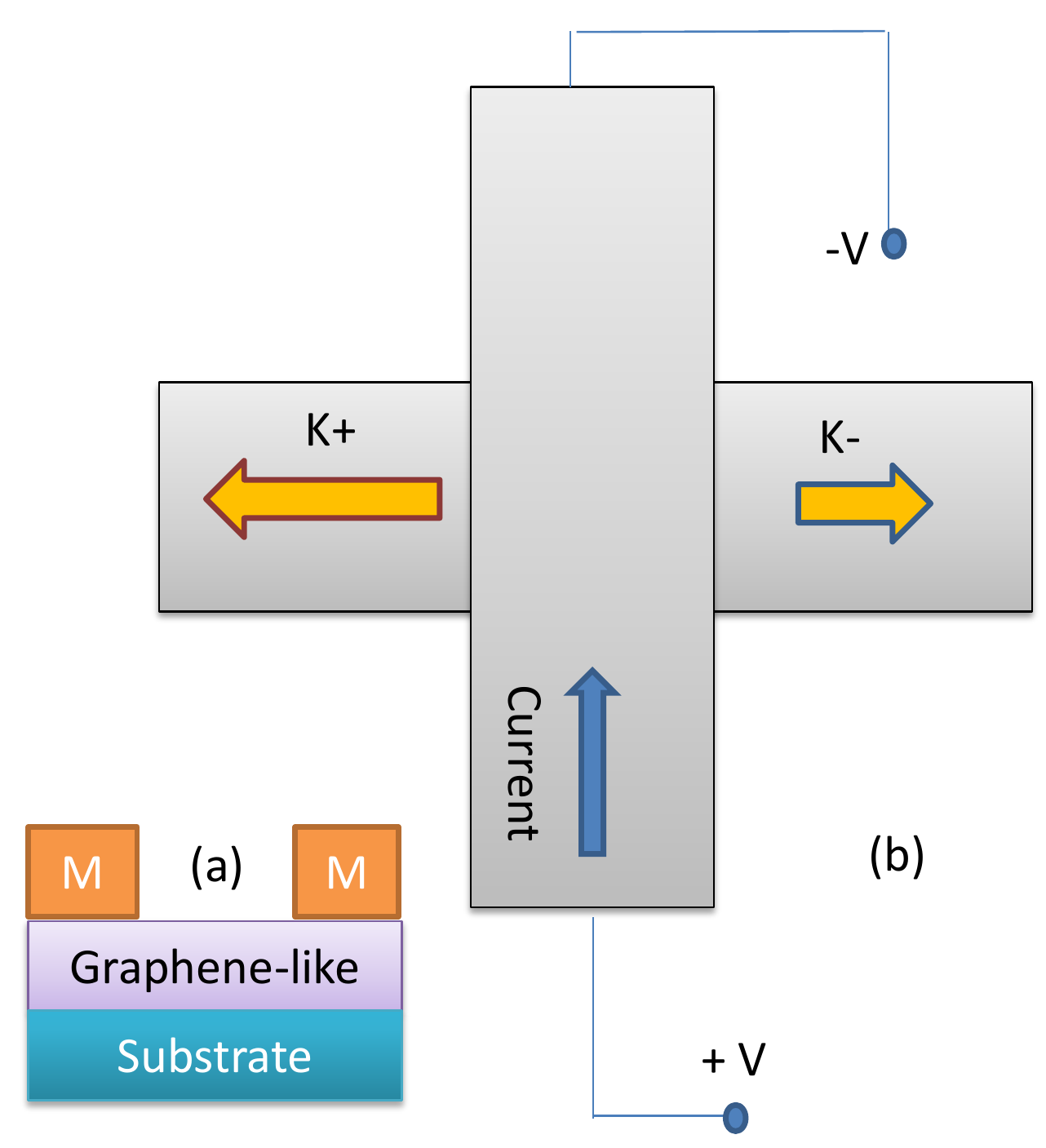}
\caption{Schematics of the realization of valley-dependent antidamping-like SOT: (a) Top view and (b) Side view. The current is injected into the vertical arm. The presence of both magnetization and staggered potential results in a nonequivalent spin density for the valleys. This leads to a different valleys on the horizontal sidearms.
}
\label{fig8}
\end{figure}

  Finally, we propose a device to detect the valley-dependent SOT. We consider a multi-terminal device as shown in Fig. \ref{fig8}. This is a typical device used to detect charge neutral-currents\cite{Gorbachev2014,Abanin2011}. The device consists of a graphene-like material sandwiched between a magnetic insulator and a non-magnetic substrate such as a topological insulator\cite{Jin}. The substrate \cite{Giovannetti} can induce a staggered potential that breaks the valley degeneracy. The voltage is applied to the sidearms and the current flows from the lower sidearm to the upper one. In the absence of magnetization, a valley Hall effect may be detected in the two horizontal terminals.\cite{Gorbachev2014} In the presence of magnetization, the torque exerted on the magnetization of the magnetic insulator deposited on top of the left or right terminal will be different.
  
\section{Conclusion}

In summary, we have investigated the nature of SOTs in two dimensional hexagonal crystals and qualitatively recovered most of the results obtained on different systems such as (Ga,Mn)As and two-dimensional Rashba gases \cite{Hang-2015}. We showed that the staggered potential and intrinsic spin-orbit coupling can strongly affect the magnitude of the torque components as well as their angular dependence. In the presence of staggered potential and exchange field, the valley degeneracy can be lifted and we obtain a valley-dependent antidamping SOT, while the field-like component remains mostly unaffected. This feature is understood in terms of Berry curvature and we show that the valley imbalance can be as high as 100\% by tuning the bias voltage or magnetization angle. 

\section*{Acknowledgement}
H.L. and A.M. were supported by the King Abdullah University of Science and Technology (KAUST).

\end{document}